\documentstyle[rauisch,epsfig]{article}
\frompage{000} \topage{000}                                              
\def\rightmark{Break-up of $~^7$Li.}
\def\leftmark{L. Fortunato}

\title{Break-up and electromagnetic response of light weakly-bound
dicluster systems.} 
\authors{
{Lorenzo Fortunato %
}\\[2.812mm]
{\normalsize
\hspace*{-8pt} Dip. di Fisica 'G.Galilei' and INFN, 
I-35100 Padova, Italy\\[0.2ex] 
}}
 
\abstract{This study is focused on the break-up
and electromagnetic response of light weakly-bound dicluster nuclei. 
The cluster picture in the case of $~^7$Li is shown to be a very good
approximation and in this framework we calculate nuclear structure
observables. We solve the Schr\"odinger equation for the relative 
motion both for discrete and continuum states 
and this automatically takes into a proper account the role of resonances.
A concentration of strength in the low energy continuum, 
solely due to the weakly-bound nature of the bound states is seen 
and explained as a favourable matching between the wavelengths of the 
initial and final states. Finally preliminary results on form factors
are briefly outlined and their microscopic derivation as well as utilization 
in reaction studies is discussed.}

\keyword{Nuclear clusters, break-up } 
\PACS{21.60.Gx, 25.60.Gc}

\begin{document}
 
\maketitle
\setcounter{page}{1}
\section{Discussion of the model}\label{1}
We adhere to the description of $~^7$Li as a dicluster system formed by an 
$\alpha$ plus a $t$. It was shown by many authors \cite{Wal} that this 
way of treating the system gives a good description of the properties 
of the nucleus under study. We solved the Schr\"odinger equation for
the relative motion of the two clusters (considered as frozen spherical
distributions of charge and mass). Both coulomb and nuclear 
interactions have been included, the former being corrected at small distances
for the finite extension of the charge distributions, the latter having
a common Woods-Saxon shape whose depth has been adjusted in each case
to give the correct energy for both bound and
resonant states in the continuum. This provides both resonant and 
non-resonant strength. The wavefunctions that we have found 
differ slightly from the ones obtained within the RGM method, 
having however very similar spatial extension.
The calculated charge and matter radii, electric and matter quadrupole 
intrinsic moments are found to be in good agreement with the 
experimental data.

\begin{figure}[t]
\vspace{-.5cm}
\begin{flushleft}
\epsfig{figure=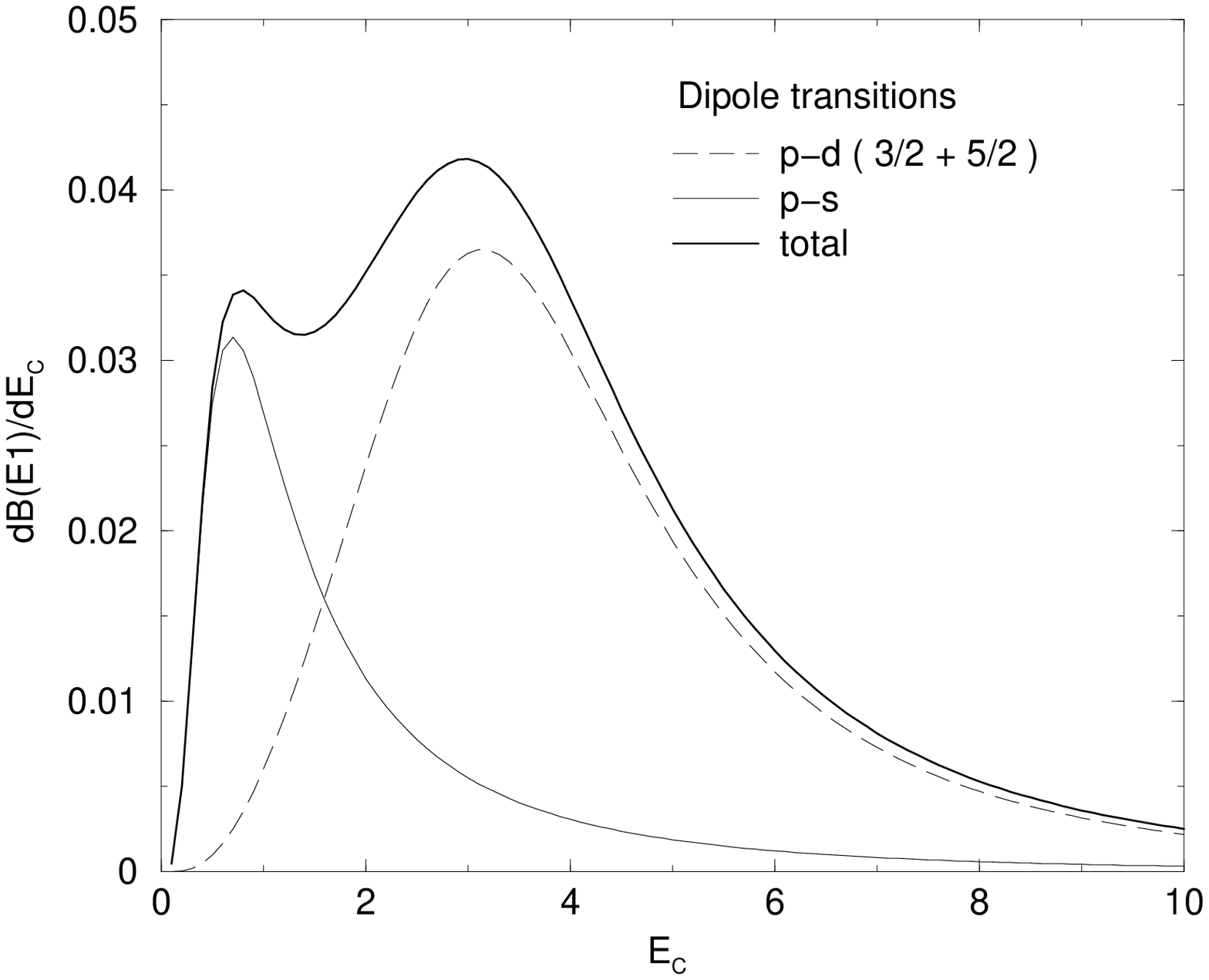,height=145pt}          
\end{flushleft}
\vspace{-164pt}
\begin{flushright}
\epsfig{figure=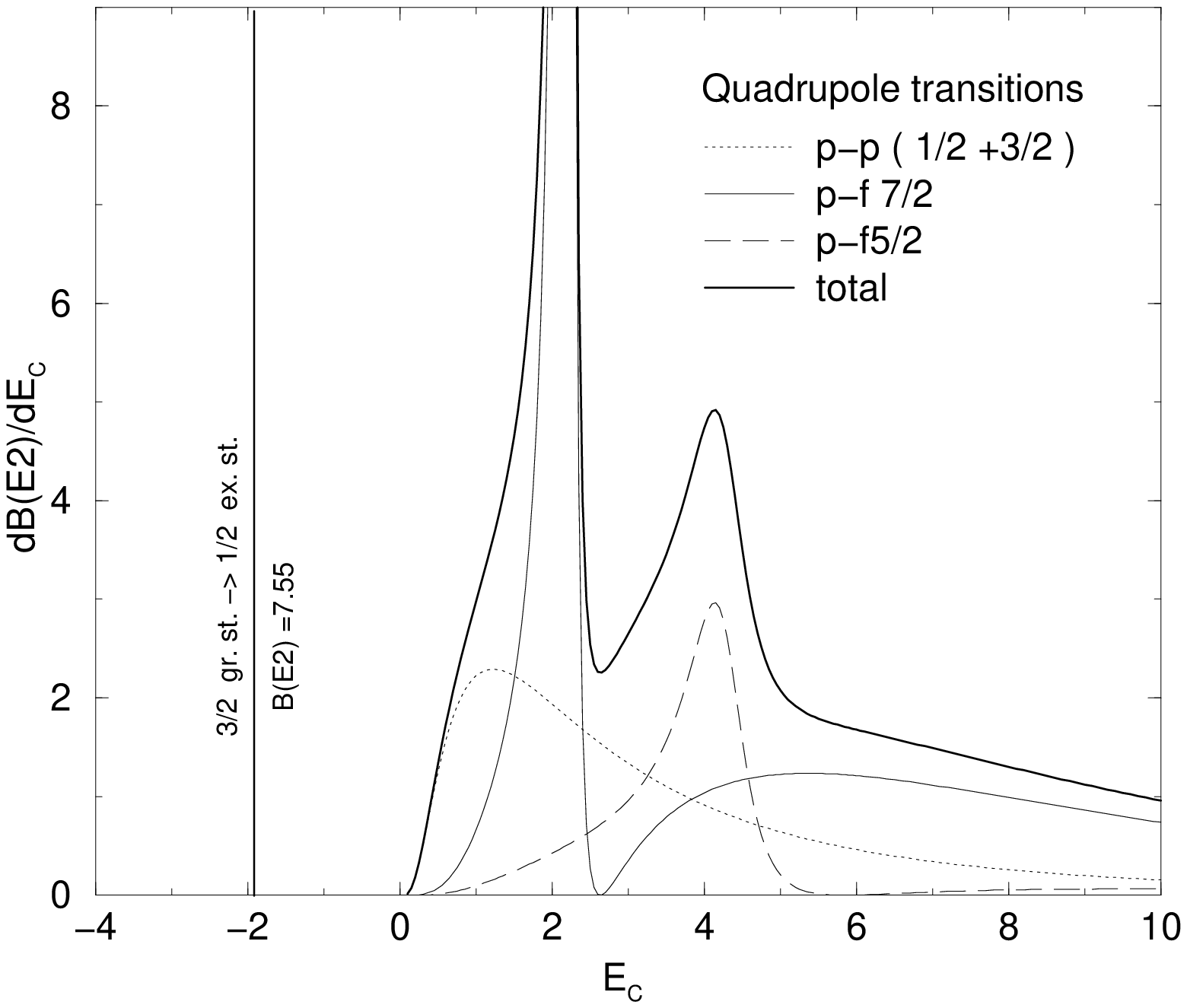,height=144pt}          
\end{flushright}
\vspace{-1.cm}
\caption[]{Electric dipole and quadrupole response (differential reduced 
transition probability with respect to energy in the continuum). 
Energies are in MeV, the zero being the threshold for break-up,
while the vertical scale is in $e^2fm^{2L}/MeV$.}
\label{fig1}\vspace{-.9cm}
\end{figure}

\section{Results}\label{2}
\vspace{-0.3cm}
We calculated the transition from the $3/2^-$ ground state to the
 $1/2^-$ excited bound state obtaining  $B(E2)=7.55 e^2fm^4$ and
$B(M1)=2.45 \mu^2$, both very close to the measured values \cite{Voe,Tok}.
The results for dipole and quadrupole transitions to the continuum are 
displayed in fig. (\ref{fig1}).
In addition to the resonant quadrupole strength associated with the f-states,
a concentration of strength at low
energy is clearly seen. It has the same nature as in \cite{Das2,Das1} 
where it is explained as due to a favourable matching between the wavelengths
of the initial weakly-bound wavefunction and of the final one 
in the continuum. \\
For what concern reaction studies
the microscopic derivation of formfactors has been
done with a twofold aim: to elucidate the dependence on the distance of two
interacting ions and on the energy in the continuum, and to set
the relevance of the nuclear break-up that
is found to be dominant still at $14 fm$ (about the double of the 
sum of radii in a reaction with a heavy target). The study of such reactions
is currently under investigation.

\vspace{-0.3cm}
\section*{Acknowledgements}
\vspace{-0.3cm}
I am particularly indebted to Andrea Vitturi for sharing with me his ideas,
intuitions and computer codes and for the patience which he had constantly 
shown towards me. The work and the 
participation to the Symposium have been both supported by {\it INFN} and 
{\it Dip. di Fisica di Padova }.

\vfill\eject

\begin{thebibliography}{999}  
\vspace{-0.3cm}
\bibitem{Wal} H.Walliser and T.Fliessbach, {\it Phys. Rev. }C{\bf 31}
(1985), 2242.
\bibitem{Voe} H.G.Voelk and D.Fick,  {\it Nucl. Phys.} A{\bf 530} (1991), 475-489.
\bibitem{Tok} Y.Tokimoto et al. {\it Phys. Rev. }C{\bf 63} (2001), 035801.
\bibitem{Das2} C.H.Dasso, S.M.Lenzi and A.Vitturi, {\it Nucl. Phys. }A{\bf 639}
(1998), 635-653 
\bibitem{Das1} C.H.Dasso, S.M.Lenzi and A.Vitturi, {\it Nucl. Phys. }A{\bf 611} 
(1996), 124-138.
\end{thebibliography}
\end{document}